\documentclass[aps,prb,twocolumn,groupedaddress,showpacs]{revtex4}

\usepackage{graphicx,amssymb,amsbsy,color}
\begin {document}

\title {An Alternative Interpretation of the Magnetic Penetration Depth
Data on Pr$_{2-x}$Ce$_x$CuO$_{4-y}$ and La$_{2-x}$Ce$_x$CuO$_{4-y}$}
\author {Khee-Kyun Voo$^{1,*}$ and Wen Chin Wu$^{2}$}
\affiliation {$^{1}$Department of Electrophysics, National Chiao
Tung University, Hsinchu 30010, Taiwan \\ $^{2}$Department of
Physics, National Taiwan Normal University, Taipei 11650, Taiwan }

\date {\today}

\begin {abstract}
We have revisited the magnetic penetration depth data on the 
electron-doped cuprates Pr$_{2-x}$Ce$_x$CuO$_{4-y}$ and
La$_{2-x}$Ce$_x$CuO$_{4-y}$. It is proposed that the transition between
nodal-gap-like and nodeless-gap-like behavior upon electron-doping [e.g.,
see M. Kim $et~al$., Phys. Rev. Lett. {\bf 91}, 87001 (2003)] can be due
to a scattering of the quasiparticles in the $d$-wave superconducting
state by an incipient or weak antiferromagnetic spin-density-wave. This
conjecture is supported by the inelastic neutron scattering and 
angle-resolved photoemission experiments on some closely related
electron-doped cuprates.
\end {abstract}

\maketitle

\newpage
\section {Introduction} 
\label {intro}

% the problem and its present situation

A common opening sentence in papers related to the high-$T_c$
cuprates is, ``After fifteen years of its discovery, the mechanism
of superconductivity in the high-$T_c$ cuprates is still under
debate...''. Meanwhile, an intimately related question, the
symmetry of the superconductivity order also still possess some
ambiguity, and nailing down the gap symmetry can lay a tread for
attempts to unveil the superconducting (SC) mechanism. Today's
situation is that, in the hole-doped ($h$-doped) compounds,
tetragonally symmetric $d_{x^2-y^2}$-wave superconductivity
($d$SC) is obtained unanimously in the underdoped and
optimally-doped compounds \cite{Van95}. In the overdoped
compounds, diverged results like fully gapped $d$+$is$, $d$+$id'$
\cite{Dag01} or orthorhombic $d_{x^2-y^2}$+$s$ \cite{YCH01} may be
obtained by different probes. For the electron-doped ($e$-doped)
compounds, observations are more converged \cite {Alf99}. A nodal SC gap,
most probably $d_{x^2-y^2}$ \cite{ALF01}, is observed at underdoped.
Upon electron-doping, a nodeless gap is developed as seen in the
magnetic penetration depth (MPD) measurement
\cite{SKL03,Kok00,Pro00,Wu93,And94,KSL03}. Moreover, the gap may
become nodal again at overdoped \cite{KSL03}. The symmetry of the
nodeless gap is controversial and various possibilities have been
proposed \cite{Mul02,KD02,DGU02}, such as a conventional $s$-wave,
or some unconventional symmetry waves like $p$+$ip'$ \cite{KYZ03}.
If one admits that the symmetry of the $e$-doped compounds could
indeed differ from its $h$-doped partner and transmutation of gap
symmetry does occur, such a situation will be bewildering. It
seems that there are different mechanisms of superconductivity
governing the $e$-doped and $h$-doped cuprates. Moreover, a
quantum phase transition (QPT) or more QPTs may be taking place in
the $e$-doped cuprates. We think that the diverging observations
of the $h$-doped cuprates at the overdoped regime are due to
difficulties in the individual experiments, but the more converged
result of the $e$-doped compounds needs to be seriously look into.

% our philosophy and insights from other e-doped compounds

More often the transition between nodal-gap and nodeless-gap
behavior in the MPD, on which most conclusions of a gap symmetry
transition are based, is ascribed to a QPT of some origin
\cite{KYZ03}. In this paper, we have chosen a more conservative
but phenomenological approach. We take two observational facts as
the input. A caveat beforehand is that those observations are not
on the compounds which the MPD data were taken, due to the fact
that a particular experimental probe usually does not access
different compounds equally well. Nevertheless, those observations
are on the $e$-doped compounds. The first observation is due to
the inelastic neutron scattering (INS) experiment, by which a
soft-frequency and momentum ${\bf q} = {\bf Q} \equiv (\pi,\pi)$
(commensurate) scattering is observed in the
Nd$_{2-x}$Ce$_x$CuO$_4$ compounds \cite{K03,YKU03,VWT04c}. The
presence of such a soft excitation in contrast to the $h$-doped
cuprates is in congruence with the fact that the antiferromagnetic
(AF) phase is more robust in the $e$-doped cuprates, and destroyed
only at higher doping levels \cite{TMK90}. The second observation
is due to the angle-resolved photoemission spectroscopy (ARPES) on
the Nd compounds. A pseudogap is seen to open up at the
intersections of the AF magnetic Brillouin zone (MBZ) boundary and
the Fermi surface (FS) \cite{ALK01}. The scattering that is
responsible for the pseudogap thus has momentum $(\pi,\pi)$ and is
sharp in momentum space. Putting the INS and ARPES observations
together, we propose that in those $e$-doped cuprates where the
MPD data were taken, there also exists a soft magnetic fluctuation
with momentum $(\pi,\pi)$ and it is coupled to the quasiparticles.

% a phenomenological model: a dSC with weak AF-SDW.

\section {Formulation and Result}
\label {form}

We will study a model system with $d$SC co-existing with a {\em
weak} AF spin-density wave (SDW). The weak SDW order is used to
model an {\em incipient} SDW, or a soft-energy, momentum $ {\bf q}
= {\bf Q}$ magnetic fluctuation. An implicit assumption in this
approximation is, all vertex corrections in the current-current
correlation function (which is related to the MPD) are neglected,
and only the self-energy which directly affects the quasiparticle
(QP) spectrum is considered. As shown by Kampf and Schrieffer
\cite{KS90}, the change in the single-particle spectrum due to a
static SDW was essentially the same as that due to the scattering
by a long coherence length, soft magnetic fluctuation (see also
Ref.~\onlinecite{KHD03}). Thus this modeling should be physically sound
for the current problem.

We start from a hamiltonian with $d$SC and AF SDW order,
\begin{eqnarray}
H = && \sum_{{\bf k} \sigma} \xi_{\bf k} c_{{\bf k}\sigma}^\dagger
c_{{\bf k}\sigma} + \sum_{\bf k} [\Delta_{\bf k} c_{{\bf
k}\uparrow}^\dagger c_{-{\bf k}\downarrow}^\dagger + {\rm H.c.}
]\nonumber\\ && + \sum_{{\bf k} \sigma}\Phi \sigma c_{{\bf
k}\sigma}^\dagger c_{{\bf k+Q}\sigma} , \label{eq:H}
\end{eqnarray}
where $\xi_{\bf k} = -2 t ( {\rm cos}~k_x + {\rm cos}~k_y ) - 4 t'
{\rm cos}~k_x ~ {\rm cos}~k_y - \mu $, $t$ is the nearest-neighbor
(NN) hopping which is set to 1, $t'=-0.4$ is the
next-nearest-neighbor (NNN) hopping used to reproduce a typical
bowed FS, ${\bf Q} \equiv (\pi,\pi) $, $\sigma = +1 ~ (-1)$ for spin
up (down), $\mu$ is the chemical potential, and $\Delta_{\bf k}$
and $\Phi$ are the SC and SDW order respectively. For a particular
band filling, a fixed $\mu$ is used at all temperatures,
superconducting and normal states. The slight change of $\mu$ is
below one percent in all our cases, and is neglected. All
equations in this paper are derived for a complex-valued
$\Delta_{\bf k}$ and a real-valued $\Phi$, but we will be
discussing only the $d_{x^2-y^2}$-symmetry SC state, i.e.,
$\Delta_{\bf k} = \Delta_{\bf k} ^* \equiv \Delta_{\rm SC} ({\rm
cos}~k_x - {\rm cos}~k_y) / 2 $. The values of $\Delta_{\rm SC}$
and $\Phi$ in our context are not solved in the usual mean field
manner, but they are $chosen$. We have chosen $\Delta_{\rm SC} =
0.1$ and $\Phi = 0.2$ \cite{VWT04b} throughout this paper, and the
resulting system is metallic (in accord with the real materials).

Defining a Nambu operator
\begin {eqnarray}
\Psi_{\bf k} \equiv \left[\matrix{c_{{\bf k}\uparrow}\cr c_{{\bf
-k}\downarrow}^\dagger \cr c_{{\bf k+Q}\uparrow} \cr c_{{\bf
-k-Q}\downarrow}^\dagger}\right], \label{eq:psi}
\end {eqnarray}
one can then define a 4$\times$4 Matsubara Green function matrix
$G({\bf k},\tau)$ component-wise by $[G({\bf
k},\tau)]_{\alpha\beta} \equiv - \langle {\rm T} _\tau \Psi_{{\bf
k}\alpha} (\tau) \Psi_{{\bf k}\beta}^\dagger \rangle$. The
temporal-Fourier-transformed SC Green function matrix $G^{\rm
SC}({\bf k},i\omega_n)$ is found to be
\begin {eqnarray}
G^{\rm SC}({\bf k},i\omega_n) = \sum_{\eta,\sigma=\pm 1} { a_{\eta
\sigma}^{\rm SC}({\bf k})  \over { i\omega_n - E_{\eta\sigma}
({\bf k}) } }, \label{eq:G}
\end {eqnarray}
where $a_{\eta\sigma}^{\rm SC}({\bf k})$ is the hermitian spectral
weight matrix whose components are
\begin {widetext}
\begin {eqnarray}
\left[ a_{\eta \sigma}^{\rm SC}({\bf k}) \right] _{11} &=& { \sigma
\over {2 [E_+({\bf k})^2-E_-({\bf k})^2]} } \left\{ \left[ 1 + { {\xi_{\bf
k}} \over { E_{\eta\sigma} ({\bf k}) } } \right]  
[ E_{\eta\sigma}({\bf k})^2-E_{\bf k+Q}^2-\Phi^2] 
+ { {\Phi^2} \over { E_{\eta\sigma} ({\bf k}) } } (\xi_{\bf k} 
+\xi_{\bf k+Q}) \right\}, \nonumber\\
\left[a_{\eta \sigma}^{\rm SC}({\bf k})\right]_{12} &=&{ {\sigma} \over {2
[E_+({\bf k})^2-E_-({\bf k})^2 ] E_{\eta\sigma}({\bf k}) } } 
\{ \Delta_{\bf k} [E_{\eta\sigma}({\bf k}) ^2 - E_{\bf k+Q}^2
] + \Delta_{\bf k+Q} \Phi^2 \},\nonumber\\ \left[a_{\eta\sigma}^{\rm
SC}({\bf k})\right] _{13} &=& { {\sigma \Phi} \over
{2 [ E_+({\bf k})^2-E_-({\bf k})^2 ] } } 
\left\{ { 1 \over {E_{\eta\sigma} ({\bf k}) } }
[ E_{\eta\sigma} ({\bf k}) ^2 - \Phi^2 + \xi_{\bf k} \xi_{\bf k+Q} 
+ \Delta_{\bf k} {\Delta_{\bf k+Q}}^\ast ] + (\xi_{\bf k}
+\xi_{\bf k+Q}) \right\}, \nonumber \\
\left[a_{\eta \sigma}^{\rm SC}({\bf k})\right] _{14} &=&{ {\sigma \Phi}
\over {2 [ E_+({\bf k})^2-E_-({\bf k})^2] E_{\eta\sigma} ({\bf k}) } }
[ E_{\eta\sigma} ({\bf k}) (\Delta_{\bf k} + \Delta_{\bf k+Q}) 
+ \xi_{\bf k} \Delta_{\bf k+Q} - \xi_{\bf k+Q} \Delta_{\bf k} ].
\label {eq:comp}
\end {eqnarray}
\end {widetext}
$[a_{\eta \sigma}^{\rm SC}({\bf k})]_{22}$, $[a_{\eta \sigma}^{\rm
SC}({\bf k})]_{23}$, and $[a_{\eta \sigma}^{\rm SC}({\bf
k})]_{24}$ are obtained from $[a_{\eta \sigma}^{\rm SC}({\bf
k})]_{11}$, $[a_{\eta \sigma}^{\rm SC}({\bf k})]_{14}$, and
$[a_{\eta \sigma}^{\rm SC}({\bf k})]_{13}$ respectively by the
substitution $\xi_{\bf k} \rightarrow -\xi_{\bf k}, \xi_{\bf k+Q}
\rightarrow -\xi_{\bf k+Q},
\Delta_{\bf k}
\rightarrow \Delta_{\bf k} ^\ast, \Delta_{\bf k+Q} \rightarrow
\Delta_{\bf k+Q} ^\ast$; while
$[a_{\eta \sigma}^{\rm SC}({\bf k})]_{33}$, $[a_{\eta \sigma}^{\rm
SC}({\bf k})]_{34}$, and $[a_{\eta \sigma}^{\rm SC}({\bf
k})]_{44}$ are obtained from $[a_{\eta \sigma}^{\rm SC}({\bf
k})]_{11}$, $[a_{\eta \sigma}^{\rm SC}({\bf k})]_{12}$, and
$[a_{\eta \sigma}^{\rm SC}({\bf k})]_{22}$ respectively by the
exchange $\xi_{\bf k} \leftrightarrow \xi_{\bf k+Q}, \Delta_{\bf k}
\leftrightarrow \Delta_{\bf k+Q}$. The poles are at $E_{\eta\sigma} ({\bf
k}) \equiv \eta E_\sigma ({\bf k})$ ($\eta, \sigma = \pm 1$), with
\begin {widetext}
\begin {eqnarray}
E_\sigma ({\bf k}) &\equiv& \sqrt { { {E_{\bf k}^2+E_{\bf k+Q}^2} \over 2
} + \Phi^2 + \sigma \sqrt { \left( { {E_{\bf k}^2-E_{\bf k+Q}^2} \over 2 }
\right) ^2  + \Phi^2[(\xi_{\bf k} + \xi_{\bf k+Q})^2+|\Delta_{\bf k} +
\Delta_{\bf k+Q}|^2 ] } }, 
\label {eq;Esigma}
\end {eqnarray}
\end {widetext}
where $E_{\bf k} \equiv ( \xi_{\bf k}^2 + |\Delta_{\bf k}|^2
)^{1/2}$. Note that for the current case $\Delta_{\bf
k+Q}=-\Delta_{\bf k}$ and the term $|\Delta_{\bf k}+\Delta_{\bf
k+Q}|$ in $E_{\sigma} ({\bf k})$ vanishes, but is kept for better
illustration. Each SC branch is splitted by the SDW or vice versa
speaking, forming four branches of excitations $E_{\eta\sigma}
({\bf k})$.

%%%%%%%%%%%%%%%%%%%%%%%%%%%%%
\begin {figure}

\includegraphics{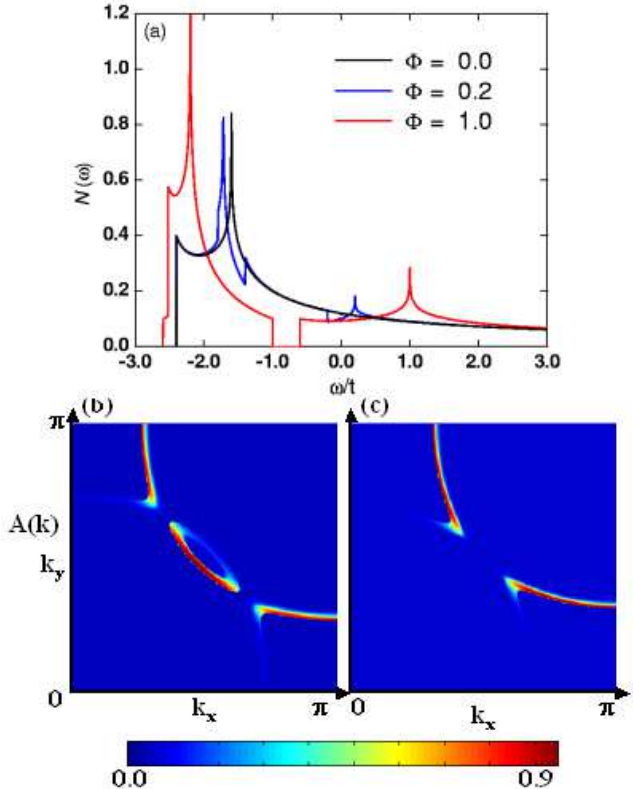}
\vspace {11cm}

\caption {
Normal state DOS with SDW order $\Phi =
$ 0 (black), 0.2 (blue), and 1 (red) are shown in (a). The
chemical potentials are set at zero. We will use $\Phi = 0.2$ for
all calculations later in this paper. The near-Fermi-level
spectral weight $A({\bf k})$ (for $\Phi = 0.2$ and $\Delta_{\rm
SC} = 0.1$) are shown for the cases (b) $\mu = -0.3$ and (c)
$+0.1$.
}
\label {fig0}
\end {figure}
%%%%%%%%%%%%%%%%%%%%%%%%%%%

% weak SDW

A few more words on what we mean by a weak SDW is desired.
Figure~\ref{fig0}(a) shows the normal state DOS for cases of
different magnitudes of the SDW order, and it is seen that for our
band with a weak SDW, no gap is formed at any chemical potential
in contrast to the strong SDW case. Nevertheless, the FS may still
be fragmented at some band fillings as seen in the distribution of
near-Fermi-level spectral weight [Fig.~\ref{fig0}(b) and (c)]. The
near-Fermi-level spectral weight as a function of momentum is
defined by $A({\bf k})\equiv \int _{-0.15|t|} ^{+0.15|t|} d\omega
~ \pi^{-1} {\rm Im} G_{11} ^{\rm SC} ({\bf k},\omega+i0) f
(\omega)$, where $f (\omega)$ is the Fermi distribution function
but here it is taken as the Heaviside step function
$\Theta(-\omega)$ for $T=0$ limit. The integration limit
``$0.15|t|$" is an arbitrarily chosen small parameter, which is
larger than $\Delta_{\rm SC}$ however.

% MPD

We now examine the low-temperature behavior of MPD. The MPD tensor
$\lambda_{\mu \nu}(T)$ is related to the zero-momentum and
zero-frequency current-current correlation function. The current
operator can be written as 
\begin {eqnarray} 
j_\mu ({\bf q}) = -e
\sum_{ {\bf k} \in {\rm MBZ} } \Psi_{\bf k}^\dagger \gamma_\mu (
{\bf k} + {{\bf q} \over 2} ) \Psi_{\bf k+q}, \label{eq:J}
\end {eqnarray}
where $\gamma_\mu({\bf k}) \equiv {\rm Diag} [{\bf
v}^\mu_{\bf k} {\bf 1}_2, {\bf v}^\mu_{\bf k+Q} {\bf 1}_2]$, ${\bf
v}^\mu \equiv {\partial \xi_{\bf k}}/{\partial {\bf k}_\mu}, {\bf
1}_2 \equiv {\rm Diag}[1,1]$ is the 2$\times$2 unit matrix, $e$ is
the electronic charge, and ${\bf k}$ is sum over the AF MBZ.
$\lambda_{\mu \nu} (T)$ is then given by
\begin {eqnarray}
&& {1 \over { \lambda _{\mu \nu} (T)^2 } } \nonumber\\  
&& = {4\pi e^2 \over {\beta N}} \sum_{{\bf k},i\omega_n}{\rm Tr} \left[
G^{\rm SC} ({\bf k}, i\omega_n) \gamma_\mu({\bf k}) G^{\rm SC} ({\bf k},
i\omega_n) \gamma_\nu({\bf k})\right. \nonumber\\ 
&& ~~~~~~~~~~~ - \left.G^{\rm N} ({\bf
k}, i\omega_n) \gamma_\mu({\bf k}) G^{\rm N}({\bf k}, i\omega_n)
\gamma_\nu({\bf k}) \right], \label{eq:p1}
\end {eqnarray}
where $G^{\rm N}$ is the normal-state Green function matrix, which
can be simply obtained from $G^{\rm SC}$ by setting $\Delta_{\rm
SC}$ to zero, but keeping the same SDW order $\Phi$. Summing off
the Matsubara frequencies, Eq.~(\ref{eq:p1}) is reduced to
\begin {eqnarray}
&& {1\over { \lambda _{\mu \nu} (T)^2 } } = {4\pi e^2 \over N}\sum_{{\bf
k}}\sum_{\eta\sigma\eta^\prime\sigma^\prime}{
{f [E_{\eta\sigma}({\bf k})] - f [E_{\eta^\prime\sigma^\prime}({\bf k}) ]}
\over
{E_{\eta\sigma} ({\bf k}) - E_{\eta^\prime\sigma^\prime} ({\bf k}) }
} \nonumber\\ && \times {\rm Tr} \left[ a_{\eta \sigma}^{\rm SC}({\bf
k}) \gamma_\mu({\bf k}) a_{\eta' \sigma'}^{\rm SC}({\bf k})
\gamma_\nu({\bf k})- (\Delta_{\rm SC} \rightarrow 0 )\right].
\label {eq:p2}
\end {eqnarray}
Due to the tetragonal symmetry, $\lambda_{\mu \nu}^{-2} = \delta_{
\mu \nu } \lambda_{\mu \mu}^{-2}$, and $\lambda_{xx} =
\lambda_{yy} \equiv \lambda$. Some further simplification can be
made and eventually this integral is evaluated numerically. The
main contribution to the integral comes from the large $ [
f(\omega)-f(\omega') ] / [ \omega-\omega' ] $, i.e., near-zero
frequency regime, and it is therefore a sum over low-energy
thermal excitations. When $\omega=\omega'$ (e.g., when
$\eta=\eta'$ and $\sigma=\sigma'$), the quantity $ [
f(\omega)-f(\omega') ] / [ \omega-\omega' ] $ is defined as the
limit $ df / d\omega $.

%%%%%%%%%%%%%%%%%%%%%%%%%%%%%
\begin {figure}

\includegraphics{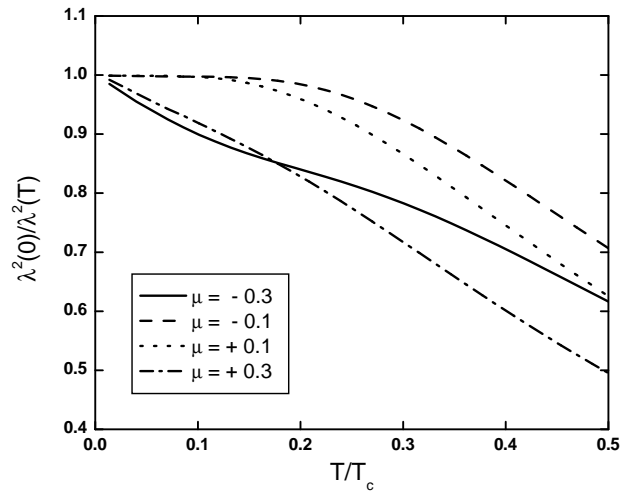}
\vspace {7cm}

\caption {
Low-temperature magnetic penetration
depth at various band fillings is plotted as $\lambda(0)^{2} /
\lambda(T)^{2}$ versus the reduced temperature $T/T_c$
($2\Delta_{\rm SC}/T_c = 4$ is used). The chemical potentials are
$\mu = -0.3, -0.1, +0.1, +0.3$ (fillings from 1.16 to 1.30 for our
band).
}
\label {fig1}
\end {figure}
% ( plot x from 0 to 0.025 (=Tc/2); y from 0.4 to 1.)
%%%%%%%%%%%%%%%%%%%%%%%

Figure \ref {fig1} shows the low-temperature MPD. The same $\Delta_{\rm
SC}$ and $\Phi$ are used within the temperature range considered, and
this approximation should be valid at least at the very low temperature
regime, say below $\sim T_c/4$. At approaching zero temperature, we find
linearly decreasing behavior $\lambda(0)^{-2} - \lambda(T)^{-2} \propto T
$ at low and high band fillings, which is essentially due to a nodal
$d$SC; and negligible temperature dependence at some intermediate band
fillings, which is due to a nodeless excitation gap in the single particle
spectrum (see Fig.~\ref {fig3}).

% speficic heat

\begin {figure}

\includegraphics{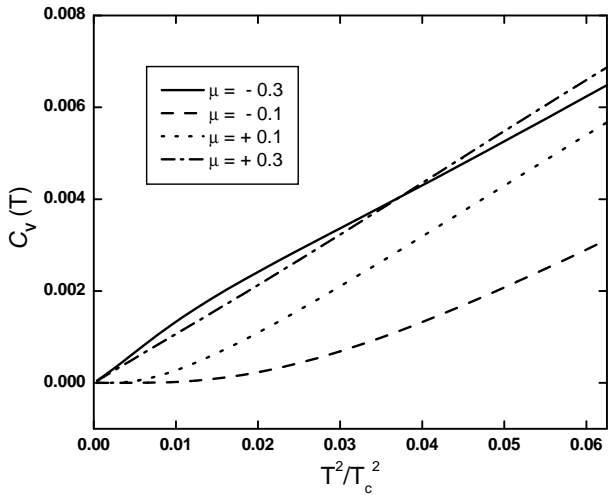}
\vspace {7cm}

\caption {
The specific heat $C_{\rm v}$ is plotted
versus $T^2$. The band parameters are the same as those in
Fig.~\ref{fig1}.
}
\label {fig2}
\end {figure}

\begin {figure}

\includegraphics{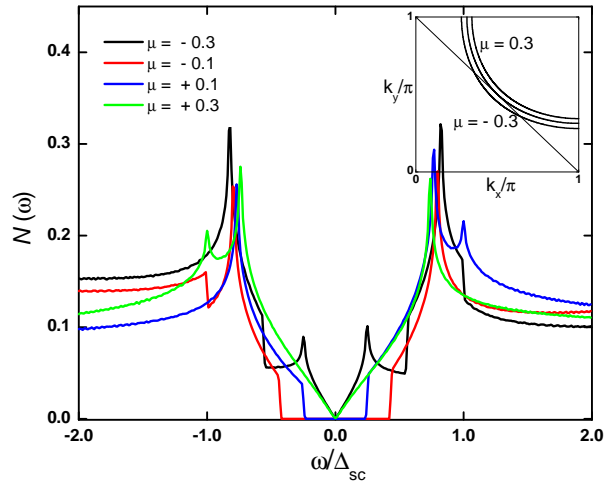}
\vspace {7cm}

\caption {
The DOS in the SC state near the Fermi
level is shown with chemical potentials $\mu = -0.3, -0.1, +0.1$,
and $+0.3$. The inset shows the Fermi surfaces when there was no
SDW order, at chemical potentials $\mu=-0.3,0,+0.3$. The MBZ
boundary is also shown.
}
\label {fig3}
\end {figure}

The transition between nodal-gap and full-gap behavior in the MPD
is compared to the results of low-temperature specific heat.
Figure~\ref {fig2} shows the constant volume specific heat
\begin {eqnarray}
C_{\rm v}(T) &=& { d \over {dT} } \sum_{ {\bf k} \in {\rm MBZ},
\eta \sigma } E_{\eta\sigma} ({\bf k}) f [E_{\eta\sigma} ({\bf
k})] \nonumber\\&=& 2 \sum_{ {\bf k} \in {\rm MBZ}, \sigma }
{E_{\sigma} ({\bf k}) ^2 \over T^2} f [ E_{\sigma} ({\bf k}) ] f
[- E_{\sigma} ({\bf k})], \label{eq:C}
\end {eqnarray}
where $ {df (\varepsilon) } / {dT} = ( \varepsilon / T ^2 )
f(\varepsilon) f(-\varepsilon) $ is used in the second line. In
congruence with the MPD, the $T^2$ dependence of $C_{\rm v} (T)$
at low temperatures can be ascribed to the linear
energy-dependence of the low-energy density of states (DOS) of the
$d$SC (i.e., $N(\omega) \propto |\omega|$ at $\omega \sim 0$, see
Fig.~\ref{fig3}); and the case of negligible temperature-dependence is
ascribable to a nodeless gap (see Fig.~\ref{fig3}).

%%%%%%%%%%%%%%%%%%%%%%%%%%%%%
% (plot $T^2$ from 0 to 1.5625E-4 [i.e., = (0.0125)^2 = (Tc/4)^2 ] ).
%%%%%%%%%%%%%%%%%%%%%%%

% DOS

Figure \ref{fig3} shows the DOS [i.e., $ N(\omega) = - (\pi N)^{-1}
\sum_{\bf k} {\rm Im} [ G^{\rm SC} ({\bf k}, \omega+i0) ] _{11}$ at
different band fillings. The full gap \cite{VWT04d} that appears at
intermediate band fillings explains the small temperature dependence of
$\lambda$ (see Fig.~\ref{fig1}) and $C_{\rm v}$ (see Fig.~\ref{fig2}) at
low temperatures.
Interestingly, a subgap twin-peak structure \cite{VWT04a} is seen
in the case of the MBZ boundary cuts the FS but the gap remain
nodal-like (the case of $\mu = -0.3$ in Fig.~\ref{fig3}). Some
``peak-dip-hump'' structure may also appear (e.g., see the
negative energy side of the $\mu=-0.1$ case in Fig.~\ref{fig3}).

%%%%%%%%%%%%%%%%%%%%%%%%%%%%%
%%%%%%%%%%%%%%%%%%%%%%%

% ARPES

\begin {figure}

\includegraphics{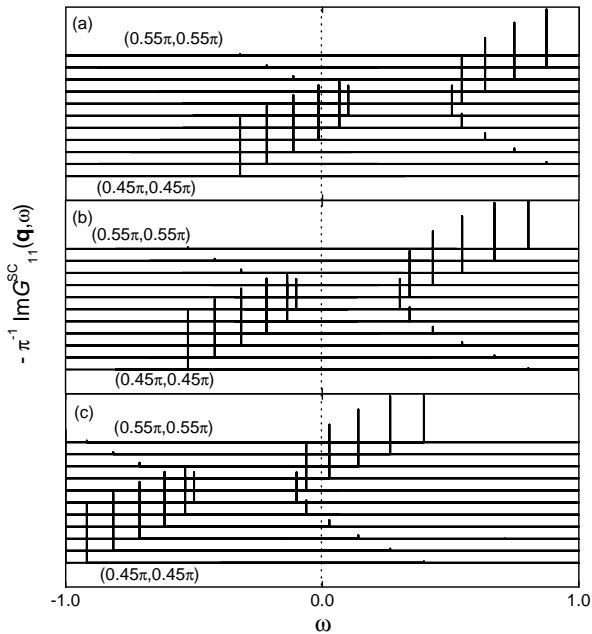}
\vspace {9cm}

\caption {
Single-particle spectrum $- \pi ^{-1} {\rm Im} ~ [ G^{\rm SC} ({\bf k},
\omega+i0) ]_{11}$ along the diagonal-${\bf k}$ direction is plotted
from  $(0.45\pi,0.45\pi)$ (lower most) to $(0.55\pi,0.55\pi)$ (upper
most), for (a) $\mu = -0.3$, (b) $\mu = -0.1$, and (c) $\mu = +0.3$.
The heights of the spikes are drawn proportional to the weight of
the poles.
}
\label {fig4}
\end {figure}

The formation of the full gap when the FS is close to the MBZ boundary is
illucidated in Fig.~\ref{fig4}. Figure~\ref{fig4} shows the 
single-particle spectrum $- \pi ^{-1} {\rm Im} [ G^{\rm SC}
({\bf k}, \omega+i0) ]_{11}$ along the diagonal-${\bf k}$
directions, where a plain $d_{x^2-y^2}$-wave SC system would have
its gapless single-particle excitations. At the presence of a SDW,
the diagonal-${\bf k}$ QP peaks do not disperse to the Fermi level
when the FS (defined at the absence of SDW) gets close to the MBZ
boundary [Fig.~\ref{fig4}(b). See also Fig.~\ref{fig0}(c)], and this
explains the full gaps in Fig.~\ref{fig3}.
Scattering by the SDW has opened up a gap at the vicinity of the
MBZ boundary. Interestingly, since the ARPES can observe only the
negative energy part of the single-particle spectrum, the QPs at
the SC nodes that are gapped by a SDW [see Fig.~\ref{fig4}(b)] may be seen
to disperse like QPs that are gapped by a SC gap. Moreover, though
the DOS may not vanish on the Fermi level, a forbidened window in
$\omega$-space at below the Fermi level may still be found for the
diagonal-${\bf k}$ QPs [see Figs.~\ref{fig3} and \ref{fig4}(c)].

%%%%%%%%%%%%%%%%%%%%%%%%%%%%%
%%%%%%%%%%%%%%%%%%%%%%%

\section {Concluding Remarks}
\label {conc}

% closing discussion

We have used the same $\Delta_{\rm SC}$, $T_c$, and $\Phi$ throughout
our discussion for simplicity. Taking into account the differences
of those parameters in various cases should make no qualitative
difference in our conclusions as we are essentially discussing thermally
activated behaviors due to a full gap, or scale-less behaviors due to a
nodal-like gap. As long as $\Phi$ is finite, there exists a range of
doping (within the realistic doping regime) that can give  rise to a full
gap. Furthermore, we also believe that our conclusion will
remain qualitatively the same when the vertex correction is taken into
account, since the gap at the MBZ boundary will always be opened up
due to the mixing of QPs at ${\bf k}$ and ${\bf k+Q}$. As the
would-be-gapless points on the FS get close to the MBZ boundary,
they are pushed to higher energies and a full gap in the total DOS is
formed.

We have adopted a phenomenological approach rather than obtaining the SDW
order from a self-consistent mean-field-theory (MFT). This is because
naive MFTs are usually insufficient to describe the very weak (probably
incipient) but persisting SDW indicated by the experiments. In such a
situation, the location of the FS but not the size of the SDW, is the
dominant factor in determining the low-energy excitations in the $d$SC
state. Note that though we have assumed a static SDW, the result is in
fact equally applicable to the case of an incipient SDW or a soft AF spin
fluctuation, as the electrons around the Fermi level are ``fast" modes,
and any relatively ``slow" spin fluctuations are essentially static.

Though the properties of the $d$SC state can be drastically modified at
some of the band fillings, the normal state properties are expected to
remain normally. For instance, the resistivity will always have a
quadratic dependence on temperature because in the normal state there
exists no gap in the DOS near the Fermi level [Fig.~\ref{fig0}(a)]. On the
impurity scattering effect on the MPD, the linear temperature dependence
of $\lambda(T)^{-2}$ at low temperatures in the clean limit that occurs
when the $d$-wave nodes on the FS are away from the MBZ boundary, is
expected to cross over to quadratic in $T$ (in the resonant scattering
limit)\cite{arberg93}, regardless of the existence of the SDW or
not. The reason for the independence of SDW is that in those cases
the low-energy nodal QPs are essentially unaffected --- a fact that can be
seen in the single-particle spectral function [see Figs.~\ref{fig4}(a) and
(c)], and manifested in the linear in $|\omega|$ dependence of the DOS at
low energies (see Fig.~\ref{fig3}). As the low-temperature MPD is related
only to the low-energy excitations of those unaffected QPs, a theory
including the effect of impurity scattering will be essentially the same
as that without SDW, which is already widely available in literature
\cite {arberg93}.

At the overdoped compounds, two possible factors might account for the
recovering of a nodal gap as seen in Ref.~\onlinecite{KSL03}. The first
is, the diagonal portions of the FS have proceeded to the other side of,
and leave the MBZ boundary (as studied in the current paper). The second
is, the strength of the magnetic fluctuation is weakened. Precise ARPES
and INS measurement on the compounds may be used to resolve the
possibilities.

We have investigated the transmutation between nodal-gap-like and
nodeless-gap-like behaviors in some of the electron-doped
cuprates. A scenario inspired by experiments on some related
compounds is proposed. We propose that there exists a weak or
incipient SDW order in the compounds, and it is coupled to the
QPs. This scattering can push the would-be-gapless diagonal QPs to
higher energies, and a $d_{x^2-y^2}$-symmetry SC system can behave
like it is nodeless.

\indent {\bf Acknowledgments -} This work is supported by the NSC
of Taiwan under Grant No. 92-2112-M-009-035 (KKV) and
92-2112-M-003-009 (WCW). We thank C.-S. Ting and L. Alff for useful
communications.


\begin {thebibliography}{99}

\bibitem[*]{coraut} To whom correspondence should be addressed. E-mail:
kkvoo@cc.nctu.edu.tw

\bibitem{Van95} D. J. van Harlingen, Rev. Mod. Phys. {\bf 67}, 515 (1995).

%\bibitem{TK00a} C. C. Tsuei and J. R. Kirtley, Phys. Rev. Lett. {\bf 85},
%182 (2000).
%\bibitem{TK00} C. C. Tsuei and J. R. Kirtley, Rev. Mod. Phys. {\bf 72},
%969 (2000).

\bibitem{Dag01} Y. Dagan $et~al.$, Phys. Rev. Lett. {\bf 87}, 177004
(2001).

\bibitem{YCH01} N.-C. Yeh $et~al.$, Phys. Rev. Lett. {\bf 87}, 87003 (2001).

%\bibitem{Alf98} L. Alff $et~al.$, J. Eur. Phys. B {\bf 5}, 423 (1998); S.
%Kashiwaya $et~al.$, Phys. Rev. B {\bf 57}, 8680 (1998).
% no andreev bound states.

\bibitem {Alf99} To our knowledge, the series of studies unveiling the
symmetry of the SC order of the $e$-doped cuprates were initiated by a
paper of L. Alff {\em et al.} in Phys. Rev. Lett. {\bf 83}, 2644 (1999).

% L. Alff et al., PRB 58, 11197 (1998).
% s-wave for el-doped compounds.

\bibitem{ALF01} N. P. Armitage $et~al.$, Phys. Rev. Lett. {\bf 86}, 1126
(2001).
% the SC order is likely d(x2-y2).

% F. Hayashi et al., J. Phys. Soc. Jpn. 67, 3234 (1998).
%
% A. Mourachkine, Europhys. Lett. 50, 663 (2000).

\bibitem{SKL03} J. A. Skinta $et~al.$, Phys. Rev. Lett. {\bf
88}, 207005 (2002).

\bibitem{Kok00} J. D. Kokales $et~al.$, Phys. Rev. Lett. {\bf 85}, 3696
(2000).

\bibitem{Pro00} R. Prozorov $et~al.$, Phys. Rev. Lett. {\bf 85}, 3700
(2000).

\bibitem{Wu93} D. H. Wu $et~al.$, Phys. Rev. Lett. {\bf 70}, 85 (1993).

\bibitem{And94} A. Andreone $et~al.$, Phys. Rev. B {\bf 49}, 6392 (1994).

\bibitem{KSL03} M.-S. Kim $et~al.$, Phys. Rev. Lett. {\bf 91}, 87001,
(2003).
% it is mentioned in the upper right column on page 3 in that paper, a
% nodeless gap appears only at intermediate dopings, whereas gaps are
% nodal in UD and OD compounds.

\bibitem{KD02} A. Kohen and G. Deutscher, cond-mat/0207382.

\bibitem{DGU02} D. Daghero $et~al.$, cond-mat/0207411.

\bibitem{Mul02} K. A. Muller, Phil. Mag. Lett. {\bf 82}, 279 (2002).

\bibitem{KYZ03} V. A. Khodel $et~al.$, cond-mat/0307454.

\bibitem{K03} H. J. Kang $et~al.$, Nature {\bf 423}, 522 (2003).

\bibitem{YKU03} K. Yamada $et~al.$, Phys. Rev. Lett. {\bf 90}, 137004
(2003).

\bibitem{VWT04c} Whereas in the $h$-doped cuprates, a commensurate INS
peak also exists but is resonance-like and at higher energies.

\bibitem{TMK90} T. R. Thurston $et~al.$, Phys. Rev. Lett. {\bf 65}, 263
(1990).

\bibitem{ALK01} N. P. Armitage $et~al.$, Phys. Rev. Lett. {\bf 87}, 147003
(2001).
% a PG exist at the MBZ boundary

\bibitem{KS90} A. P. Kampf and J. R. Schrieffer, Phys. Rev. B {\bf 42},
7967 (1990).

\bibitem{KHD03} B. Kyung $et~al.$, cond-mat/0312499.

\bibitem{VWT04b} In the sense of the usual mean-field theory,
$\Phi = U \langle S_z \rangle$, where $U$ is the onsite Hubbard
repulsion and $S_z$ is the $z$-component of the spin operator.
Therefore $\Phi = 0.2$ and $U \sim 5$ implies $\langle S_z \rangle
\sim 0.04$.

\bibitem{VWT04d} Note that this full gap is not simply the SDW gap, as
it is obviously much smaller than the SDW order $\Phi = 0.2$ chosen in
this paper. It is due to the intertwined $d$SC and SDW order.

\bibitem{VWT04a} Such subgap structure in the DOS resembles that of
a $d_{x^2-y^2}$$+$$s$ gap (e.g., see Ref.~\onlinecite{YCH01}) or some
sign-changing extended-$s$ gap [e.g., see G. Zhao, Phys. Rev. B
{\bf 64}, 24503 (2001)], which are essentially sign-changing gaps
with more than one absolute magnitudes of gap
maxima.

\bibitem{arberg93} See, e.g., P. Arberg, M. Mansor, and J. P. Carbotte, J.
Phys. Chem. Solids {\bf 54}, 1461 (1993).

%\bibitem{BFQ03} A. Biswas $et~al.$, Phys. Rev. Lett. {\bf 88}, 207004
%(2002).

%\bibitem{A02} N. P. Armitage $et~al.$, Phys. Rev. Lett. {\bf 88}, 257001
%(2002).
% what is this paper talking about?

%\bibitem{Zha01} G. Zhao, Phys. Rev. B {\bf 64}, 24503 (2001).

\end {thebibliography}
\end{document}